\documentclass[twoside,twocolumn,9pt]{article}
\usepackage{extsizes}
\usepackage[super,sort&compress,comma]{natbib} 
\usepackage[version=3]{mhchem}
\usepackage[left=1.5cm, right=1.5cm, top=1.785cm, bottom=2.0cm]{geometry}
\usepackage{balance}
\usepackage{times,mathptmx}
\usepackage{sectsty}
\usepackage{textcomp}
\usepackage{graphicx} 
\usepackage{lastpage}
\usepackage[format=plain,justification=raggedright,singlelinecheck=false,font={stretch=1.125,small,sf},labelfont=bf,labelsep=space]{caption}
\usepackage{float}
\usepackage{fancyhdr}
\usepackage{fnpos}
\usepackage[english]{babel}
\usepackage{array}
\usepackage{droidsans}
\usepackage{charter}
\usepackage[T1]{fontenc}
\usepackage[usenames,dvipsnames]{xcolor}
\usepackage{setspace}
\usepackage[compact]{titlesec}

\newcommand{\fig}{Fig.}

\newcommand{\figref}[1]{\fig~\ref{#1}}

\newcommand{\tabref}[1]{table~\ref{#1}}

\renewcommand{\eqref}[1]{equation~(\ref{#1})}

\newcommand{\vecnp}[1]{\mathbf{#1}}

\usepackage{soul}
 \renewcommand{\hl}[1]{#1}

\usepackage{epstopdf}

\definecolor{cream}{RGB}{222,217,201}

\begin{document}

\pagestyle{fancy}
\thispagestyle{plain}
\fancypagestyle{plain}{

\fancyhead[C]{\includegraphics[width=18.5cm]{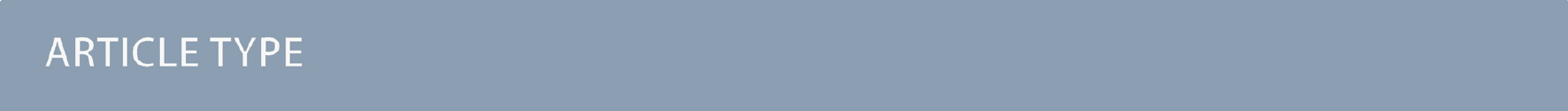}}
\fancyhead[L]{\hspace{0cm}\vspace{1.5cm}\includegraphics[height=30pt]{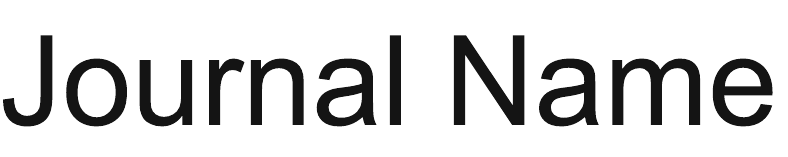}}
\fancyhead[R]{\hspace{0cm}\vspace{1.7cm}\includegraphics[height=55pt]{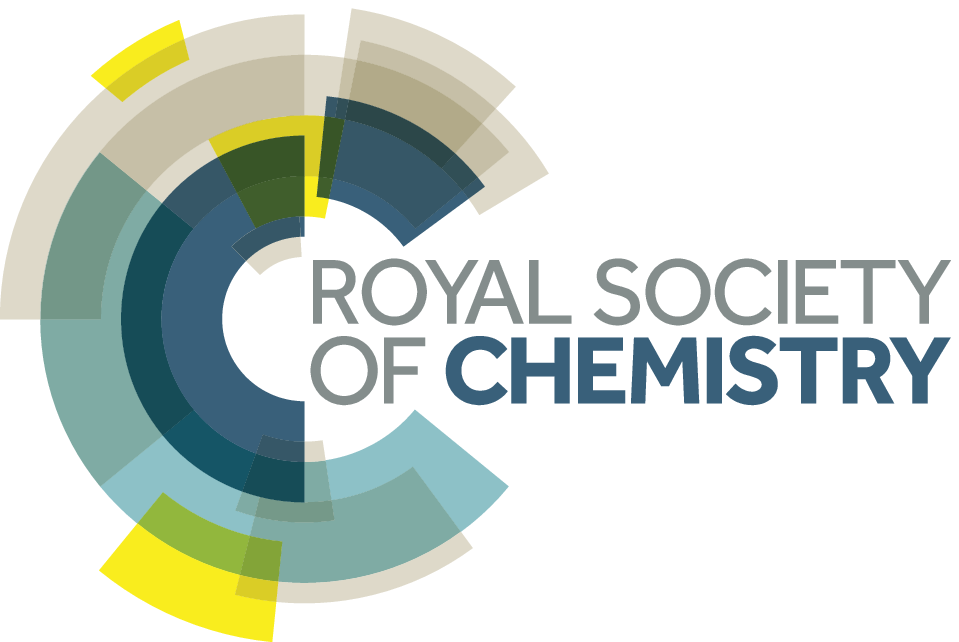}}
\renewcommand{\headrulewidth}{0pt}
}

\makeFNbottom
\makeatletter
\renewcommand\LARGE{\@setfontsize\LARGE{15pt}{17}}
\renewcommand\Large{\@setfontsize\Large{12pt}{14}}
\renewcommand\large{\@setfontsize\large{10pt}{12}}
\renewcommand\footnotesize{\@setfontsize\footnotesize{7pt}{10}}
\makeatother

\renewcommand{\thefootnote}{\fnsymbol{footnote}}
\renewcommand\footnoterule{\vspace*{1pt}%
\color{cream}\hrule width 3.5in height 0.4pt \color{black}\vspace*{5pt}} 
\setcounter{secnumdepth}{5}

\makeatletter 
\renewcommand\@biblabel[1]{#1}            
\renewcommand\@makefntext[1]%
{\noindent\makebox[0pt][r]{\@thefnmark\,}#1}
\makeatother 
\renewcommand{\figurename}{\small{Fig.}~}
\sectionfont{\sffamily\Large}
\subsectionfont{\normalsize}
\subsubsectionfont{\bf}
\setstretch{1.125} 
\setlength{\skip\footins}{0.8cm}
\setlength{\footnotesep}{0.25cm}
\setlength{\jot}{10pt}
\titlespacing*{\section}{0pt}{4pt}{4pt}
\titlespacing*{\subsection}{0pt}{15pt}{1pt}

\fancyfoot{}
\fancyfoot[LO,RE]{\vspace{-7.1pt}\includegraphics[height=9pt]{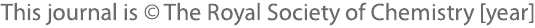}}
\fancyfoot[CO]{\vspace{-7.1pt}\hspace{13.2cm}\includegraphics{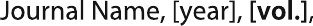}}
\fancyfoot[CE]{\vspace{-7.2pt}\hspace{-14.2cm}\includegraphics{head_foot/RF}}
\fancyfoot[RO]{\footnotesize{\sffamily{1--\pageref{LastPage} ~\textbar  \hspace{2pt}\thepage}}}
\fancyfoot[LE]{\footnotesize{\sffamily{\thepage~\textbar\hspace{3.45cm} 1--\pageref{LastPage}}}}
\fancyhead{}
\renewcommand{\headrulewidth}{0pt} 
\renewcommand{\footrulewidth}{0pt}
\setlength{\arrayrulewidth}{1pt}
\setlength{\columnsep}{6.5mm}
\setlength\bibsep{1pt}

\makeatletter 
\newlength{\figrulesep} 
\setlength{\figrulesep}{0.5\textfloatsep} 

\newcommand{\topfigrule}{\vspace*{-1pt}%
\noindent{\color{cream}\rule[-\figrulesep]{\columnwidth}{1.5pt}} }

\newcommand{\botfigrule}{\vspace*{-2pt}%
\noindent{\color{cream}\rule[\figrulesep]{\columnwidth}{1.5pt}} }

\newcommand{\dblfigrule}{\vspace*{-1pt}%
\noindent{\color{cream}\rule[-\figrulesep]{\textwidth}{1.5pt}} }

\makeatother

\twocolumn[
  \begin{@twocolumnfalse}
\vspace{3cm}
\sffamily
\begin{tabular}{m{4.5cm} p{13.5cm} }

\includegraphics{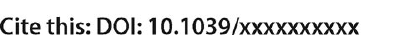} & \noindent\LARGE{\textbf{Comparison of Classical Reaction Paths and Tunneling Paths studied with the Semiclassical Instanton Theory$^\dag$}} \\
\vspace{0.3cm} & \vspace{0.3cm} \\

 & \noindent\large{Jan Meisner,\textit{$^{a}$} Max N. Markmeyer,\textit{$^{a}$} Matthias U. Bohner,\textit{$^{a}$} and Johannes K\"{a}stner$^{\ast}$\textit{$^{a}$}  } \\

\includegraphics{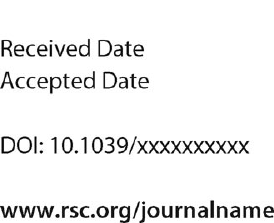} &
\noindent\normalsize{Atom tunneling in the hydrogen atom transfer reaction of the
2,4,6-tri-\emph{tert}-butyl\-phenyl radical to 3,5-di-\emph{tert}-butylneophyl,
which has a short but strongly curved reaction path, was investigated using
instanton theory. We found the tunneling path to deviate qualitatively from
the classical intrinsic reaction coordinate, the steepest-descent path in
mass-weighted Cartesian coordinates. To perform that comparison, we
implemented a new variant of the predictor-corrector algorithm for the
calculation of the intrinsic reaction coordinate.
\hl{
We used the reaction force analysis method as a mean to decompose the reaction barrier into structural and electronic components
structural and electronical components.
}
Due to the narrow energy
barrier atom tunneling is important in the abovementioned reaction, even above room
temperature.
Our calculated rate constants between 350~K and 100~K agree well
with experimental values. We found a H/D kinetic isotope effect of almost
$10^6$ at 100~K. Tunneling dominates the protium transfer below 400~K and the
deuterium transfer below 300~K.
\hl{
We compared the lengths of the tunneling path and the classical path 
for the hydrogen atom transfer in the reaction HCl + Cl 
and quantified the corner cutting in this reaction.
At low temperature, the tunneling path is about 40\% shorter than the classical path.
}
}
\end{tabular}

 \end{@twocolumnfalse} \vspace{0.6cm}

  ]

\renewcommand*\rmdefault{bch}\normalfont\upshape
\rmfamily
\section*{}
\vspace{-1cm}


\footnotetext{\textit{$^{a}$~Institute for Theoretical Chemistry, University of Stuttgart, Pfaffenwaldring 55, Stuttgart, Germany. Fax: +49-(0)711-685 64442; Tel: +49-(0)711-685 64473; E-mail: kaestner@theochem.uni-stuttgart.de}}

\footnotetext{\dag~Electronic Supplementary Information (ESI) available: Full description of the implemented algorithm, results of the functional benchmark, and reaction rate constants. See DOI: 10.1039/b000000x/}




\section{Introduction}

Atom tunneling plays an important role in chemistry as it enhances the
reaction rate constants of chemical reactions.  At very low temperatures it
determines stability and reactivity.\cite{sch11a,koz14,koz15a} The
tunneling of atoms, in particular hydrogen atoms, is important in different
fields of chemistry ranging from biochemistry \cite{koh03,lay14,var15} to
astrochemistry.\cite{ham13,bro14} Several reviews about atom tunneling were
published recently.\cite{miy04,kae14,mei16,borden2016}

There are many methods to compute the effect of atom tunneling on the rate
constants of chemical reactions.\cite{pu06,nym14,kae14} In principle, a full
quantum mechanical description of the nuclear wave function is a fully
rigorous treatment.\cite{gar95,mar11} These methods require the solution of
the time-dependent Schr\"{o}dinger equation, which poses huge computational
demands if the problem exeeds a few atoms.

A much simpler approach of including the quantum mechanical tunneling effect
is to use classical rate constants and correct them by tunneling through
approximate potential functions for which the tunneling probability can be
calculated analytically. The most prominent of these approaches use
rectangular barriers, parabolic barriers\cite{bel80} or Eckart
barriers.\cite{eck30} These approaches assume the tunneling particles to take
the same path as particles crossing the potential energy barrier classically.
They are sometimes referred to as one-dimensional tunneling corrections.

The tunneling probability, however, is increased by shortening the tunneling
path. While the average classical path, \emph{i.e.,} the minimum energy path (MEP), proceeds through a first-order saddle point on the potential energy surface (PES), the transition structure (TS), the tunnelling path cuts the corner on the concave side of curved reaction
paths on the expense of higher potential energy. \cite{mar77,tru03,fer07}
Methods which take this into account are sometimes termed
multidimensional tunneling corrections,
like the small curvature tunneling correction (SCT).\cite{sko81}
Such methods still rely on the classical MEP as a
reference for the tunneling path of the particles.

In this work we use the semiclassical instanton
theory\cite{lan67,lan69,mil75,col77,cal77,gil77,aff81,
col88,han90,ben94,mes95,ric09,ric11,alt11,ric16,mcconnell2017} based on Feynman's path integral
formalism.\cite{fey48}
At temperatures below the crossover temperature
\begin{equation}
 T_\text{c}  = \frac{\hbar \omega_\text{TS}}{2 \pi k_\text{B}}.
 \label{eq:2}
\end{equation}
atom tunneling dominates the reaction rate and instanton theory is applicable.
The underlying idea is to optimize a tunneling path,
the instanton, for each temperature by making the Euclidean action stationary.
\hl{Instanton
theory has become a useful method to study reactions and is nowadays an
established approach to calculate rate constants in different fields of
chemistry.\mbox{ \cite{cha75,mil94,mil95,mil97,sie99, sme03,qia07,and09,
  gou10a,gou10,jon10,gou11,gou11b, rom11,rom11b,mei11,gou11a,ein11,rom12,
  kry12,kae13,alv14,kry14,mei16,alv16,son16,lam16,lam17}}
The instanton is particularly advantageous when the tunneling path
qualitatively deviates from the classical path
for example at very low temperatures.\mbox{\cite{alv14,ric16a}}
In these cases 
instanton theory was found to be superior 
to MEP based methodologies like SCT.\mbox{\cite{and09}}
}

\begin{figure}[h!]
  \begin{center}
    \includegraphics[width=8cm]{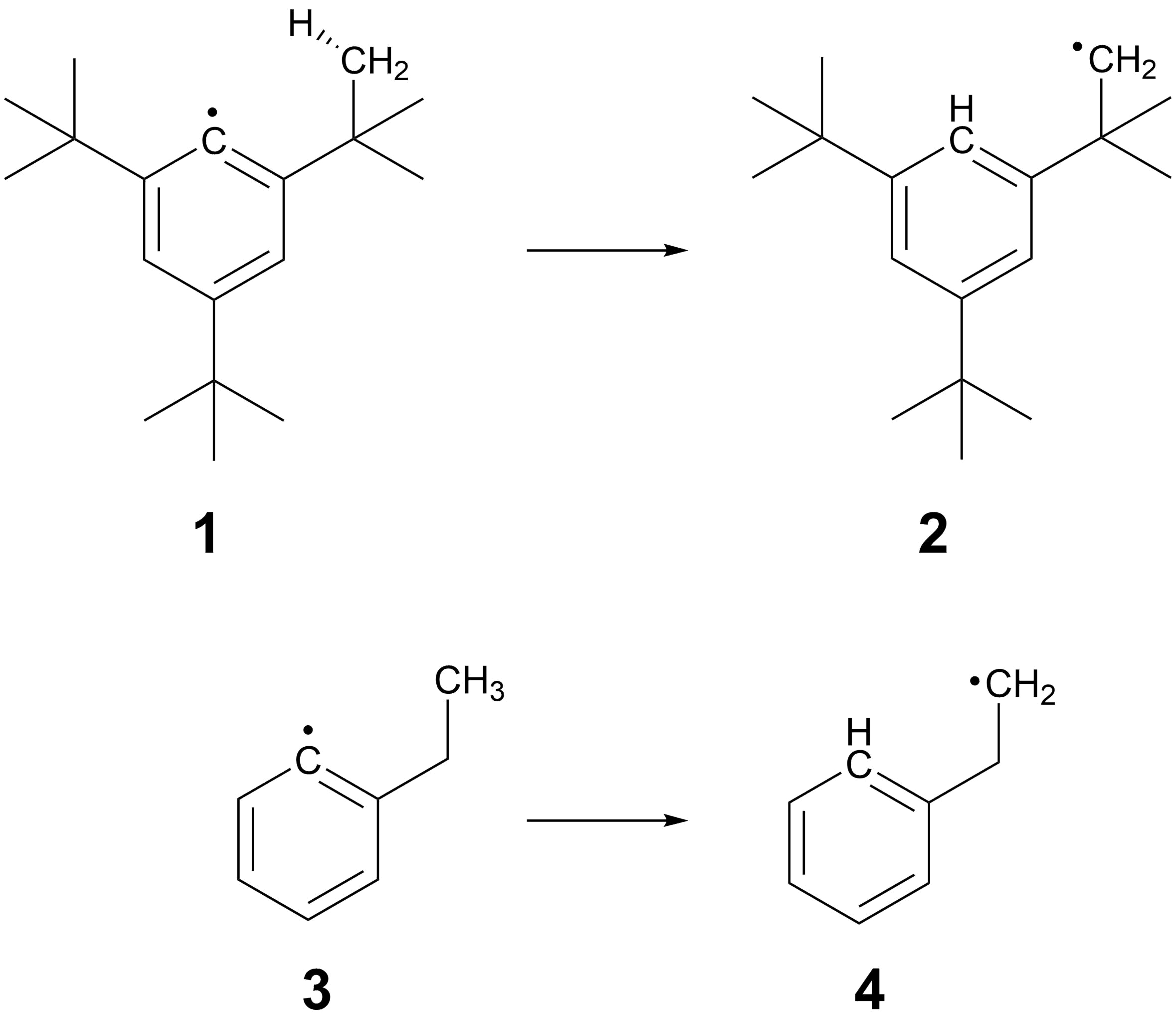}
    \caption{Reactions discussed in this study.
      \label{fig:reaction}
    }
  \end{center}
\end{figure}

The isomerization of aryl radicals as described by Brunton \emph{et
  al.}\cite{bru76} is one example of hydrogen atom transfer reactions 
where atom tunneling has a pronounced effect on the reactivity.
Brunton \emph{et al.} studied various rate
constants by means of electron paramagnetic resonance spectroscopy pointing
out that quantum mechanical tunneling is the reason for the strong
non-Arrhenius behavior.\cite{bru76}  The isomerization of the 2,4,6-tri-\emph{tert}-butylphenyl
radical ({\bf 1}) to the 3,5-di-\emph{tert}-butylneophyl radical ({\bf 2}), see
\figref{fig:reaction}, was studied from 113~K to 247~K.  The authors also
performed experiments where they substituted all methyl hydrogen atoms by
deuterium, which is here referred to as the perdeuterated system.  For that,
rate constants from 123~K to 293~K were measured.  At the lowest temperatures
studied, Brunton \emph{et al.} found a H/D kinetic isotope effect (KIE) larger than four
orders of magnitude due to atom tunneling.\cite{bru76}

In this paper we study the reaction {\bf 1} $\rightarrow$ {\bf 2} using
instanton theory. Particular emphasis is put on the reaction path with and
without tunneling. The most likely classical reaction path is the intrinsic
reaction coordinate (IRC), which is the mass-weighted MEP in Cartesian coordinates.
To compare the IRC to the instanton path, we implemented IRC search algorithms in our DL-FIND code.\cite{kae09a}
We present a modified version of the
Hessian predictor-corrector (HPC) algorithm by Hratchian \emph{et
  al.}\cite{hra04,hra05} to
determine IRCs.
\hl{Besides  {\bf 1} $\rightarrow$ {\bf 2}, we use the reaction HCl +
  Cl to compare the lengths of the
classical reaction path and instanton tunneling paths. 
}

\section{Methods}

\subsection{Intrinsic Reaction Coordinate}

The IRC connects the reactant's and product's minima and allows insight into the
mechanism of chemical reactions. It is defined as the steepest descent MEP
in mass-weighted Cartesian coordinates
$\vecnp{x}$.\cite{fuk81} The evaluation of the IRC begins at a saddle point of
first order, the transition structure, and follows the negative
of  $\vecnp{g(x)}$, the gradient of the multidimensional PES:
\begin{equation}
  \frac{\rm d\vecnp x}{{\rm d}s} = - \frac{\vecnp{g(x)}}{|\vecnp{g(x)}|}
  \label{eq:irc}
\end{equation}
Here $s$ is the arc length along the MEP in mass-weighted Cartesian coordinates.

We implemented a variant of the well-established
Hessian predictor-corrector (HPC) method by Hratchian \emph{et al.}\cite{hra04}
In summary, the idea is to use a fast integration method (here we use the explicit Euler integration) as a first estimation, the so-called predictor step.
After that, a more sophisticated method is used to improve this first
estimation, called corrector step. In the HPC integrator, the latter is determined by means of a modified Bulirsch--Stoer (mBS) integrator as described by Hratchian \emph{et al.}\cite{hra04}
The original Bulirsch--Stoer algorithm is described well elsewhere. \cite{NumericalRecipes}

For the predictor step,
the original HPC approach\cite{hra04} solves the integration of \eqref{eq:irc} analytically,
which is possible when using local quadratic approximation (LQA).\cite{pag88,pag90}
This requires a matrix diagonalization rendering the treatment of bigger systems difficult.\cite{hra10}
To avoid the diagonalization, the predictor step of the HPC was changed
to a plain explicit Euler integration 
resulting in the Euler-predictor-corrector (EulerPC) presented by Hratchian \emph{et al.}:\cite{hra10,hra11}
\begin{equation}
\vecnp{x}_{i+1}^{[P]} = \vecnp{x}_i - \Delta s  \frac{\vecnp{g(x}_i)}{|\vecnp{g(x}_i)|}
\end{equation}
In our implementation, we build a Taylor series
up to quadratic order, \emph{i.e.,} including the
Hessian matrix
and use the simple explicit Euler method  for the predictor step
to avoid the matrix diagnolization,
yet improving the quality of the predictor step by incorporation
of quadratic information.

Other aspects of the HPC were treated as described in the literature.\cite{hra04}
The full description of the technical implementation into DL-Find
including the first step and termination
is sketched in the Supplementary Information.

\subsection{Computational Details}

To give reliable rate constants, the underlying electronic
potential has to be accurate. Wave-function-based correlation methods
like CCSD(T)-F12 provide a good
solution of the electronic Schr\"{o}dinger equation but are not
suitable for the direct use in our study because of their computational
effort, especially for the calculation of gradients and Hessians of
the potential energy, as required by instanton theory.
Therefore, and because of the large number of function evaluations necessary
for the optimization of stationary points and IRCs on the PES as well as instantons, we decided to use
density functional theory
(DFT). 
Finding a functional which describes the 
reaction adequately in comparison to reliable correlation
methods was still too costly in this case.
Therefore we used a reduced model,
namely the isomerization of phenylethyl radical {\bf 3} to
ethylenebenzene radical {\bf 4}.  This reaction is very similar to the
reaction of {\bf 1} to {\bf 2} and can serve as a benchmark.

Initially, geometries were optimized using the B3LYP density
functional \cite{dir29,sla51,vosko1980,bec88,lee88,bec93} and the
def2-SVP\cite{wei05} basis set.  On these geometries, energies were
computed using explicitly correlated unrestricted coupled-cluster with
singles and doubles excitations including perturbative treatment of
triple excitations (CCSD(T)-F12)\cite{adl07,kni09a} based on a
restricted Hartree--Fock (RHF) reference function and the
cc-pVTZ-F12\cite{pet08} basis set.  The two relevant energy
differences -- the electronic activation energy $E_\text{A}$ and the
electronic reaction energy $\Delta E$ were then compared to the
corresponding values obtained by single point energy calculations with
different functional/basis set combinations.  For that, we applied
commonly used density functionals
(B3LYP\cite{dir29,sla51,vosko1980,bec88,lee88,bec93}
PBE\cite{dir29,sla51,per92,per96}
PBE0\cite{dir29,sla51,per92,per96,perdew1996_pbe0}
BP-86\cite{dir29,sla51,vosko1980,bec88,perdew1986}
BHLYP,\cite{dir29,sla51,vosko1980,bec88,lee88,becke1993A}
TPSS\cite{dir29,sla51,per92,tao2003}
TPSSH\cite{dir29,sla51,per92,tao2003,staroverov2003} M06
\cite{zhao2008} ) and the basis sets def2-SVP\cite{wei05},
def2-TZVP\cite{wei05}, and def2-TZVPD\cite{rap10}.  We also tested the
influence of a D3 dispersion correction.\cite{grimme2010}


The CCSD(T)-F12 calculations were carried out in
Molpro\cite{werner2012} version 2012.1 with the
cc-pVTZ-F12\cite{pet08} basis set.

DFT energies, gradients and second derivatives were calculated in the Turbomole program package version 7.0.1.\cite{TURBOMOLE}
SCF energies were iterated until the energy of two successive
iterations changes by less than $1.0 \cdot 10^{-9}$ a.u. on the
\emph{m5} multigrid.\cite{eichkorn1997}
First and second derivatives with respect to the nuclear coordinates (gradients and Hessians) are calculated analytically.

All geometry optimizations, IRCs, instantons and rate constants
have been calculated with DL-Find \cite{kae09a} interfaced to ChemShell.\cite{she03,met14}
The IRC path is calculated with a step size of $\Delta s = 0.04$
mass-weighted atomic units.
Hessian updates according to Bofill's formula \cite{bofill1994} were
used throughout the whole calculation of the IRC.\cite{hra05}
Stationary points were identified based on the number of imaginary
frequencies: zero for minimum structures and exactly one for the
transition structures.

Instantons and rate constants
were calculated
using sequential cooling:
the instanton at a particular temperature is used as a starting guess for the next lower temperature
and the Hessians are used for a quasi-Newton--Raphson optimizer.\cite{rom11,rom11b}
The Feynman path was discretized to 40 images down to 214~K and to 78 images down to 100~K.
The convergence with respect to the number of images was shown by an additional
instanton calculation using 154 images at 100~K
where the rate constant deviated by less than 0.2~\% from the rate constants obtained with 78 images.
All coordinates (135 degrees of freedom in case of {\bf 1}$\rightarrow$ {\bf 2}) were optimized during the instanton search until the maximum component of the gradient
was less than $1.0 \cdot 10^{-8}$ a.u. (1 a.u.~$ =a_0\sqrt{m_\text{e}}$).

\hl{
For the reaction HCl + Cl we are 
interested in the shape of the classical reaction path compared to
instanton paths.
For that we have chosen the B3LYP functional\mbox{\cite{dir29,sla51,vosko1980,bec88,lee88,bec93}} and the def2-SVP basis set\mbox{\cite{wei05}} due to their computational efficiency 
and assume that geometries, reaction paths, and instantons are reasonably represented.
}

\section{Results}

In this section
we describe the electronic potential energy of the isomerization of {\bf 1} to {\bf 2}
including a benchmark of the different density functionals using the reduced model reaction
{\bf 3}$\rightarrow$ {\bf 4}.
Following that, the IRC of the reaction obtained with the newly implemented algorithm
is discussed.
\hl{%
We present rate constants, compare them to literature data,\mbox{\cite{bru76}}
and analyze the tunneling path.
Finally, we discuss the reaction of HCl + Cl and quantify 
the corner cutting effect.
}

\subsection{Electronic Structure}
Eight commonly used density functionals with and without D3 correction for
dispersion were tested against CCSD(T)-F12/cc-pVTZ-F12 energies on
B3LYP/def2-SVP geometries.  CCSD(T) is sometimes referred to as the gold
standard of quantum chemistry as long as the electronic structure of the
studied molecule can be expected to be a single reference case.  The
explicitly correlated variant of it, named CCSD(T)-F12, improves the basis set
convergence of the electronic energy such that a triple-$\zeta$ basis can be
assumed to be sufficiently close to the basis set limit due to improved
convergence of the correlation energy.\cite{adl07,kni09a}

A legitimation for the
assumption that the reaction has just minor multireference character is given by
the T1 and D1 diagnostics which are
T1~$ = 0.013$ and D1~$ = 0.043$ for the reactant,
T1~$ = 0.011$ and D1~$ = 0.030$ for the product, and
T1~$ = 0.012$ and D1~$ = 0.030$ for the TS.
They are thus below the threshold  of
T1~$ = 0.045$ and D1~$ = 0.050$ for open shell systems.\cite{janssen1998,lambert2006}

The CCSD(T)-F12 calculations resulted in an electronic activation energy of
$E_\text{A}=86.72$~kJ~mol$^{-1}$ and a electronic reaction energy of $\Delta E
=-41.46$~kJ~mol$^{-1}$.

The numerical results of the functional benchmark are shown in the
Supplementary Information.
For the accurate calculation of rate constants, the region around
the transition structure is most important, \emph{i.e.,}  the electronic energy
barrier $E_\text{A}$ has to fit the reference CCSD(T)-F12 values.  The
B3LYP-D3 method provides the smallest deviation and underestimates the barrier
by just 3.43~kJ~mol$^{-1}$. Almost all functionals describe the electronic
reaction energy $\Delta E$ nicely, nearly independently of the basis set.  The
error of 2.35~kJ~mol$^{-1}$ for the electronic reaction energy of the
B3LYP-D3/def2-TZVP combination is acceptable. For this reaction, there is no
need for diffuse functions as for all functionals, the error introduced by
neglecting them is smaller than 1~kJ~mol$^{-1}$. Overall we have chosen the
B3LYP functional with the D3 dispersion correction and the def2-TZVP basis set
as an appropriate method leading to a reliable electronic potential for the
reaction of {\bf 3} to {\bf 4} and
we assume that the  reaction
{\bf 1} $\rightarrow$  {\bf 2} is also well described by this method.

\subsection{Intrinsic Reaction Coordinate}

Using B3LYP-D3/def2-TZVP the electronic activation energy is $E_\text{A} =
78.9$~kJ~mol$^{-1}$ and the electronic reaction energy is $\Delta E = -31.6
$~kJ~mol$^{-1}$.  Inclusion of the harmonically approximated vibrational
zero-point energy (ZPE) changes these numbers slightly to $E_\text{A,ZPE}
=64.0$~kJ~mol$^{-1}$ and $\Delta E_\text{ZPE} =-35.8$~kJ~mol$^{-1}$.  The imaginary
frequency at the transition structure is $1811i$~cm$^{-1}$ which leads to a
crossover temperature of $ T_\text{c} = 414$~K.

During the reaction, the hydrogen atom migrates from a tert-butyl group
to the phenyl ring to form a C--H-$\sigma$ bond with the aryl carbon atom.
Thermodynamically, the reaction is favored because of the high instability of the aryl radical {\bf 1}.
During the reaction and in particular at the TS,
the two carbon atoms of the tert-butyl group, two carbon atoms
of the phenyl ring, and the transferred hydrogen atom form a
\hl{planar}
five-membered ring which reduces the distance the hydrogen atom has to cover.\cite{bru76}

\begin{figure}[bt]
  \includegraphics[width=8cm]{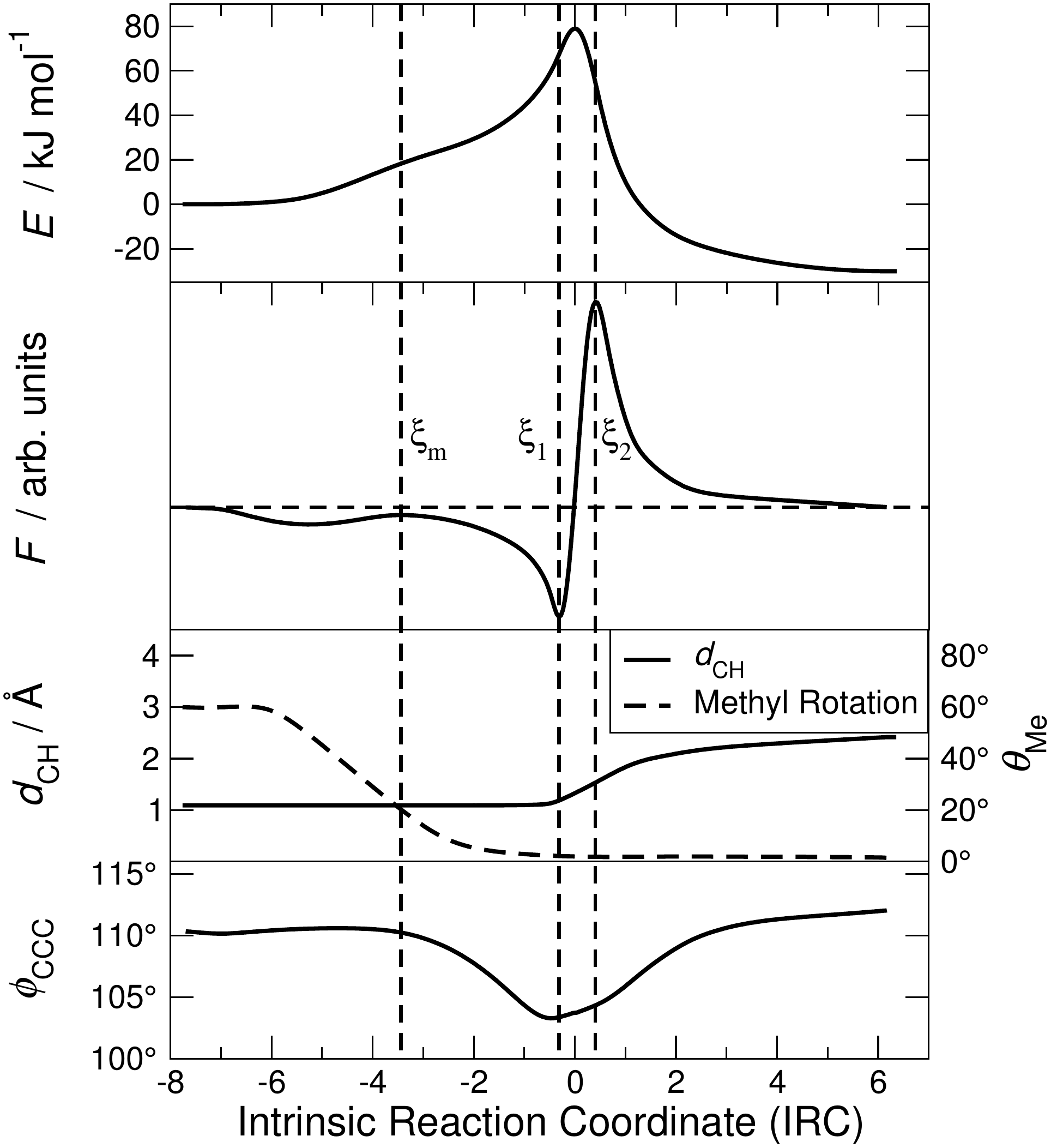}
  \caption{
    \hl{
      Geometrical and energetical characterization of the progress from  
      {\bf 1} to {\bf 2}. 
      The vertical dashed lines correspond to the characteristic points 
      of the reaction force mentioned in the text.
      First graph: Potential energy along the IRC for the hydrogen transfer
      reaction of {\bf 1} $\rightarrow$ {\bf 2}. 
      Second graph: Reaction force $F$ during the reaction.
      The two lowest graphs show 
      the  bond distance $d_\text{CH}$ of the broken C--H bond 
      and $\theta_{\text{Me}}$, the angle of the methyl rotation
      and $\phi_{\text{CCC}}$, the C--C--C angle of three
      of the carbon atoms involved in the five-membered ring in the transition
      structure, respectively.
    }
    \label{fig:irc}
  }
\end{figure}

\begin{figure}[bt]
  \begin{center}
  \includegraphics[width=8cm]{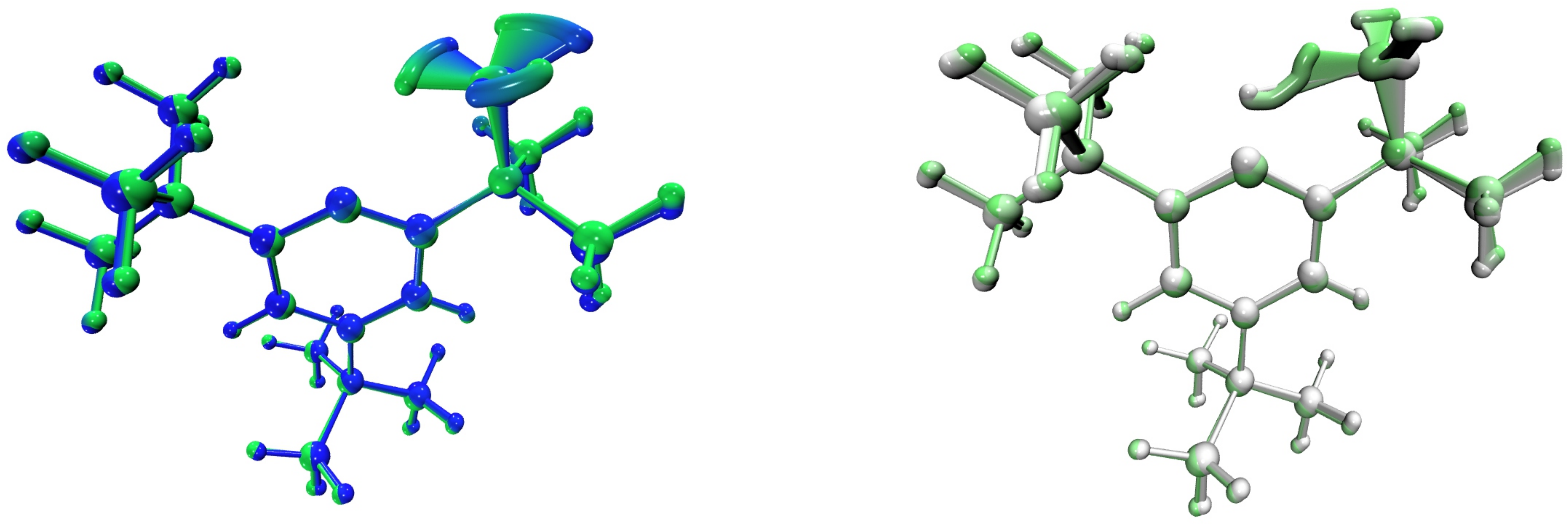}
    \caption{Geometrical representation of the IRC.
             The IRC can be split into two parts:
             the rotation of the methyl group (in the left picture from blue
             (\textbf{1}) to green)
             and the C--H bond breaking (in the right picture from green to
             white (\textbf{2})).
            \label{fig:IRC-color}
            }
  \end{center}
\end{figure}

We calculated the IRC and the corresponding potential energy, see \figref{fig:irc}.
In the direct vicinity of the TS, the curvature of the potential energy along
the barrier is astonishingly high.  Note that the distance in mass-weighted
Cartesian coordinates is very short.  This is also shown by the high absolute
value of the imaginary frequency of $1811$~cm$^{-1}$.  On the product side of
the reaction profile, the system directly proceeds down the potential energy
surface towards the product structure.  On the reactant side of the barrier,
the gradient of the potential energy with respect to the IRC diminishes and a
shoulder arises in the plot of the potential energy against the reaction
coordinate as visible in the top graph in \figref{fig:irc}.

\begin{figure}[h!tbp]
  \begin{center}
    \includegraphics[width=8cm]{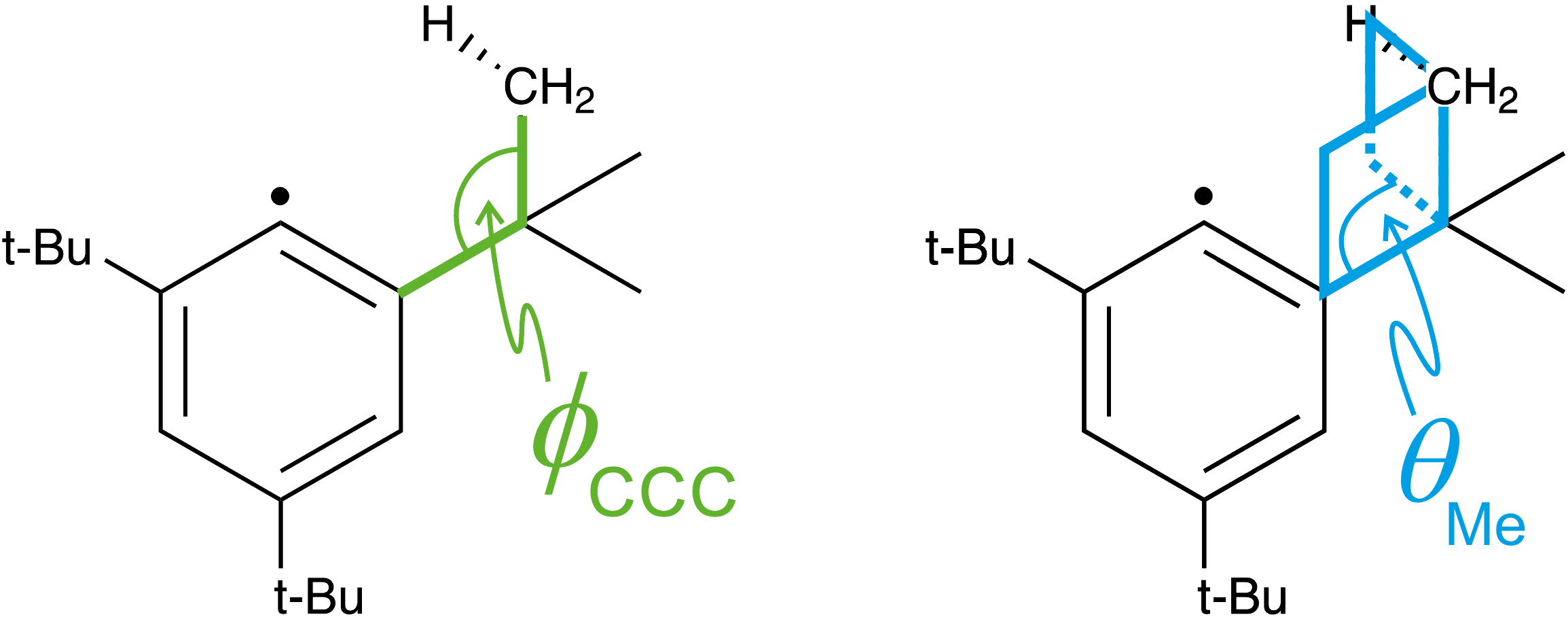}
    \caption{Definition of the angles $\phi_\text{CCC}$ \hl{(left)}
		and $\theta_{\text{Me}}$ \hl{(right)} 
		 in \textbf{1}.
            \label{fig:angles}
            }
  \end{center}
\end{figure}

To study the progress of the reaction along the IRC we concentrated on three
internal coordinates, while all degrees of freedom were included in the IRC search: $d_{\text{CH}}$, the distance of the C--H bond to be
broken during the reaction, $\theta_{\text{Me}}$, a C--C--C--H torsion angle
describing the rotation of the methyl group losing the hydrogen atom around
the aliphatic C--C bond, and $\phi_\text{CCC}$, the angle of three carbon atoms
involved in the five-membered ring of the transition structure.  The two angles
are indicated in \figref{fig:angles}.  In the reactant \textbf{1} these are
$d_{\text{CH}}= 1.092$~\AA, $\theta_{\text{Me}}\approx 60.0 ^{\circ}$, and
$\phi_\text{CCC}=110.4^{\circ}$. The coordinates of the product \textbf{2} are
$d_{\text{CH}}= 2.416$~\AA, $\theta_{\text{Me}}\approx 1.0 ^{\circ}$, and
$\phi_\text{CCC}=112.0^{\circ}$.  The deviation of $\theta_{\text{Me}}$ from zero is
due to numerical inaccuracies.

The initial step in the reaction is a rotation of the methyl group described
by $\theta_{\text{Me}}$ in order to bring the hydrogen atom closer to the
phenyl radical. This change in $\theta_{\text{Me}}$, while
$d_{\text{CH}}$ is almost unchanged, is nicely visible in the middle panel of
\figref{fig:irc}. It causes the shoulder in the energy along the IRC.  The
energy is further increased by intramolecular distortion expressed in the
change in $\phi_\text{CCC}$. From around $\phi_\text{CCC}=110^{\circ}$ in
\textbf{1} it is reduced to $103^{\circ}$ close to the TS and relaxes back to
$112^{\circ}$ in \textbf{2}. This distortion in $\phi_\text{CCC}$ is necessary in
order to bring the transferred hydrogen atom closer to the phenyl radical.
The breaking of the C--H bond finally increases the energy sharply until the
TS is reached. After that, the system relaxes directly to \textbf{2}.

\hl{
In order to gain more insight into the chemical reaction 
\textbf{1} $\rightarrow$ \textbf{2}
we calculated the reaction force, i.e., 
the negative derivative of the potential energy along the IRC with respect to
the path length:\mbox{\cite{toro-labbe1999,jaque2000}}
}
\begin{equation}
F = - \frac{\text{d} V(\xi)}{\text{d} \xi  } 
  \label{eq:force}
\end{equation}
\hl{%
The reaction force analysis is an interpretation to distinguish between 
structural and electronic effects during the course of a chemical reaction\mbox{\cite{ortega-moo2017}}.
This can be achieved
by dividing the reaction into three regions 
which are separated at the extremal points of the reaction force:
In the region from the reactant structure to $\xi_1$
structural and conformational changes cause an increase of potential energy. 
In the regions from $\xi_1$ to $\xi_{\text{TS}}$ and from $\xi_{\text{TS}}$ to  $\xi_2$ the part of the 
potential energy barrier caused by electronic effects and the 
potential energy obtained due to the formation of the new bond can be determined, respectively.
In the region from $\xi_2$ to the product structure, 
the relaxation of the molecular structure leads to a release of potential energy.
The structural and electronic contributions to the activation barrier can be quantized by $W_1$ and $W_2$,
and the structural and electronic contributions to the release of energy after passing the transition state can be quantized by $W_3$ and $W_4$, respectively:\mbox{\cite{gutierrez-oliva2005,ortega-moo2017}}
}
\begin{align*}
  &W_1= - \int_{\xi_\text{RS}}^{\xi_{1}} F(\xi) \text{d}\xi		\hfill
  &W_2= - \int_{\xi_{1}}^{\xi_\text{TS}} F(\xi) \text{d}\xi		\\
  &W_3= - \int_{\xi_\text{TS}}^{\xi_{2}} F(\xi) \text{d}\xi		\hfill
  &W_4= - \int_{\xi_{2}}^{\xi_\text{PS}}  F(\xi) \text{d}\xi
  \label{eq:force2}
\end{align*}
\hl{%
where $\xi_\text{RS}$, $\xi_\text{TS}$, and $\xi_\text{PS}$
are the position of the reactant structure, transition structure, and product structure on the reaction path, respectively.
The results can be seen in \mbox{\figref{fig:irc}}, second graph, where vertical lines indicate the separation of the three different regions.
}


\hl{
In the reaction from \textbf{1} to  \textbf{2}
the contribution of the structural changes to the potential energy barrier, $W_1$,
is higher than the contributions caused by the electonic changes, $W_2$, see \mbox{\tabref{tab:works}}.
The strong structural distortion of the carbon backbone, 
as can be seen in the change of $\phi_\text{CCC}$ close to the transition structure,
see \mbox{\figref{fig:irc}}, is the source of the huge barrier height.
A reduction of this structural stress could therefore lower the potential activation barrier.
The impact of this structural change on atom-tunneling though, would then have to be re-evaluated, of course.
The structural distortion could be reduced when including a further methylene (CH$_2$) group
leading to the trineopentylphenyl radical.
This was already tested by Brunton \emph{et al.}: they reported that the trineopentylphenyl radical
was not observed even at $-160$\textcelsius{}. Thus, they concluded that the following six-membered 
ring has the optimal spatial arrangement.

The reaction force profile in \mbox{\figref{fig:irc}} shows a minimum 
at around $\xi = -5.28$ displaying the conformational change,
\emph{i.e.,} the rotation of the methyl group.%
\mbox{\cite{yepes2012}} 
After the maximum at  $\xi = -3.44$ the 
structural distortion of the C--C--C angle $\phi_\text{CCC}$ takes place.
We can therefore separate the work of structural distortion neccessary for the reaction, $W_1$ 
into the contribution of the methyl rotation and the distortion of the carbon backbone.
We call these contributions $W_{1,\text{Me}}$ and  $W_{1,\text{struc}}$, 
see \mbox{\tabref{tab:works}}. 
For the reaction energy, the structural relaxation work, $W_4$, is larger than the electronic
work. The same trends can be seen, namely 
}
\begin{equation}
| W_3  | < | W_4|,
\end{equation}
\hl{%
indicating that the release of energy due to conformational change is larger than the release of energy due to 
the formation of the C--H bond.

In summary, the large potential energy barrier stems from structural
distortions. Therefore, atom tunneling is 
facilitated by structural strain and not, as one could intuitively assume, by a high electronic contribution to the barrier.
}

\begin{table}[h!]
 \caption{
   \hl{
     Structural and electronic contributions during the course of the reaction
     \textbf{1} $\rightarrow$ \textbf{2}.
     $W_1$ is separated into the structural work caused by the methyl
     rotation, $W_{1,\text{Me}}$ and
     the remaining structural distortion of, \emph{e.g.}, the binding angle $\phi_\text{CCC}$.
     All values in kJ~mol$^{-1}$.
   }
   \label{tab:works}
 }
 \begin{center}
\begin{tabular}{rrrrrrrrr}
\hline
 $W_{1,\text{Me}}$& $ W_{1,\text{struc}}$	 	& W$_2$ & W$_3$  &W$_4$ 	\\
18.3 & 49.0  & 11.7  & $-$24.1  &$-$85.0  \\
\hline
\end{tabular}
\end{center}
\end{table}

\subsection{Instanton Calculations}

The most likely tunneling path, \emph{i.e.,} the instanton, has to be optimized for each
temperature. Changes in the mass, as in the calculation of KIEs, also require
a re-optimization of the instanton. In principle, for secondary KIEs
(KIEs which come from substituting other atoms than the transferred hydrogen atom by
their heavier isotopes)
the tunneling path can be approximated to remain unchanged.\cite{mei11} The spread
of the instanton can be interpreted as the delocalization of the individual
atoms involved in the chemical process. Lighter atoms tend to be more
delocalized.  Since the geometrical shape of the instanton qualitatively
changes with temperature, we first discuss that before reporting on rate
constants.

\subsubsection{Tunneling Path}

As mentioned above, the instanton path can deviate from the classical reaction
path (the IRC), especially at lower temperatures.  To aid the discussion, the
shapes of the instantons and the IRC are projected onto the two variables
$d_\text{CH}$ and $\theta_{\text{Me}}$ chosen in the last section, while both IRC and instantons were always obtaind by optimizing the full coordinate set.

\begin{figure}[h!tbp]
  \begin{center}
    \includegraphics[width=8cm]{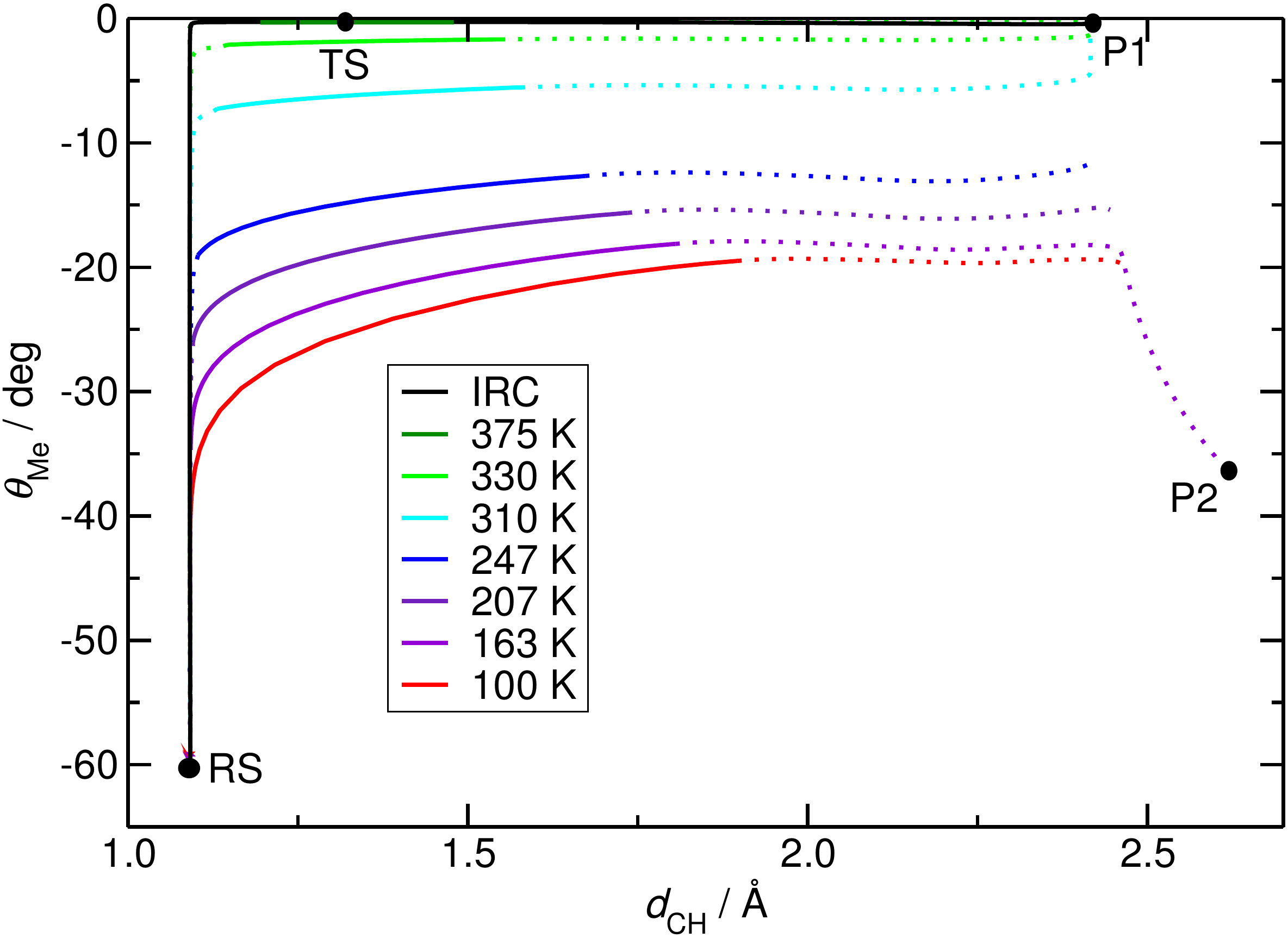}
  \caption{Shape of instantons and IRC projected onto $\theta_{\text{Me}}$ and $d_{\text{CH}}$.
           $C_s$ symmetry corresponds to $\theta_{\text{Me}}=0 ^{\circ}$ and $\theta_{\text{Me}}=60 ^{\circ}$.
           The dotted lines correspond to steepest-descent paths starting from the
           instantons' end points.
           \label{fig:instanton_coords}
          }
  \end{center}
\end{figure}

The IRC per definition starts from the reactant structure (\textbf{RS} =
minimum-geometry of \textbf{1}), proceeds via the {\bf TS} and ends on the
product side at {\bf P1}.  Both {\bf TS} and {\bf P1} have
$\theta_{\text{Me}}=0^{\circ}$ and are, therefore $C_s$ symmetric (\emph{i.e.,}
have a mirror plane). \textbf{RS} is also $C_s$ symmetric even though
$\theta_{\text{Me}}=60^{\circ}$.

High-temperature instantons are close to the IRC, $\theta_{\text{Me}}\approx
0^{\circ}$. The instanton at 375~K in \figref{fig:instanton_coords} therefore
coincides with part of the IRC. The instantons are shorter than the IRC,
however, because they only connect the points in configuration space, which are
located between the classical turning points at a given temperature. Note that the turning points, in general, do not lie on the IRC.
We connected them to the minimum geometries in \figref{fig:instanton_coords} by
calculating the steepest-descent paths in mass-weighted coordinates starting
from the endpoints of the instantons using the same algorithm as for calculating the IRC.
These connections are shown as dotted lines in
\figref{fig:instanton_coords}.

Below a temperature of 350~K the instantons start to deviate from the $C_s$
symmetry and a qualitative corner cutting effect is found. Similar phenomena
were observed for other systems
previously.\cite{mar77,tru03,fer07,mei11,alv16} The continuations by
steepest-descent paths connect these instantons to the RS geometry and, for
the instantons above 300~K, to the product geometry {\bf P1}.  At even lower
temperature, the elongation  of the tunneling path to the
product's side of the barrier ends up in a minimum structure {\bf P2}, which is asymmetric with
respect to the methyl rotation angle $\theta_{\text{Me}}$.

Both structures {\bf P1} and {\bf P2} were identified as minima by frequency
analyses. {\bf P2} is lower than {\bf P1} by
1.5~kJ~mol$^{-1}$. A nudged-elastic band calculation\cite{jon98,hen00b,hen00a,gou09a} showed a
potential energy barrier of merely 0.73~kJ~mol$^{-1}$ between them.  It can
therefore be assumed that {\bf P1} and {\bf P2} interchange even at
temperatures as low as 100~K with a rate much higher than that of
{\textbf{1} $\rightarrow$ \textbf{2}}.

\hl{
Here it has to be mentioned that the IRC is strongly curved 
(see \mbox{\figref{fig:instanton_coords}}) 
and the instantons are qualitatively different from the MEP due to corner cutting.
At the lowest temperatures presented in this work, the instanton reaction rates even lead
to a transition to another minimum structure, which could not have been detected with MEP based 
tunneling methods.
The optimization of the correct tunneling path is therefore necessary for the correct description 
of the reaction rate constants which will be presented in the following section.
}

\subsubsection{Rate Constants}

\begin{figure}[htbp]
  \begin{center}
    \includegraphics[width=8cm]{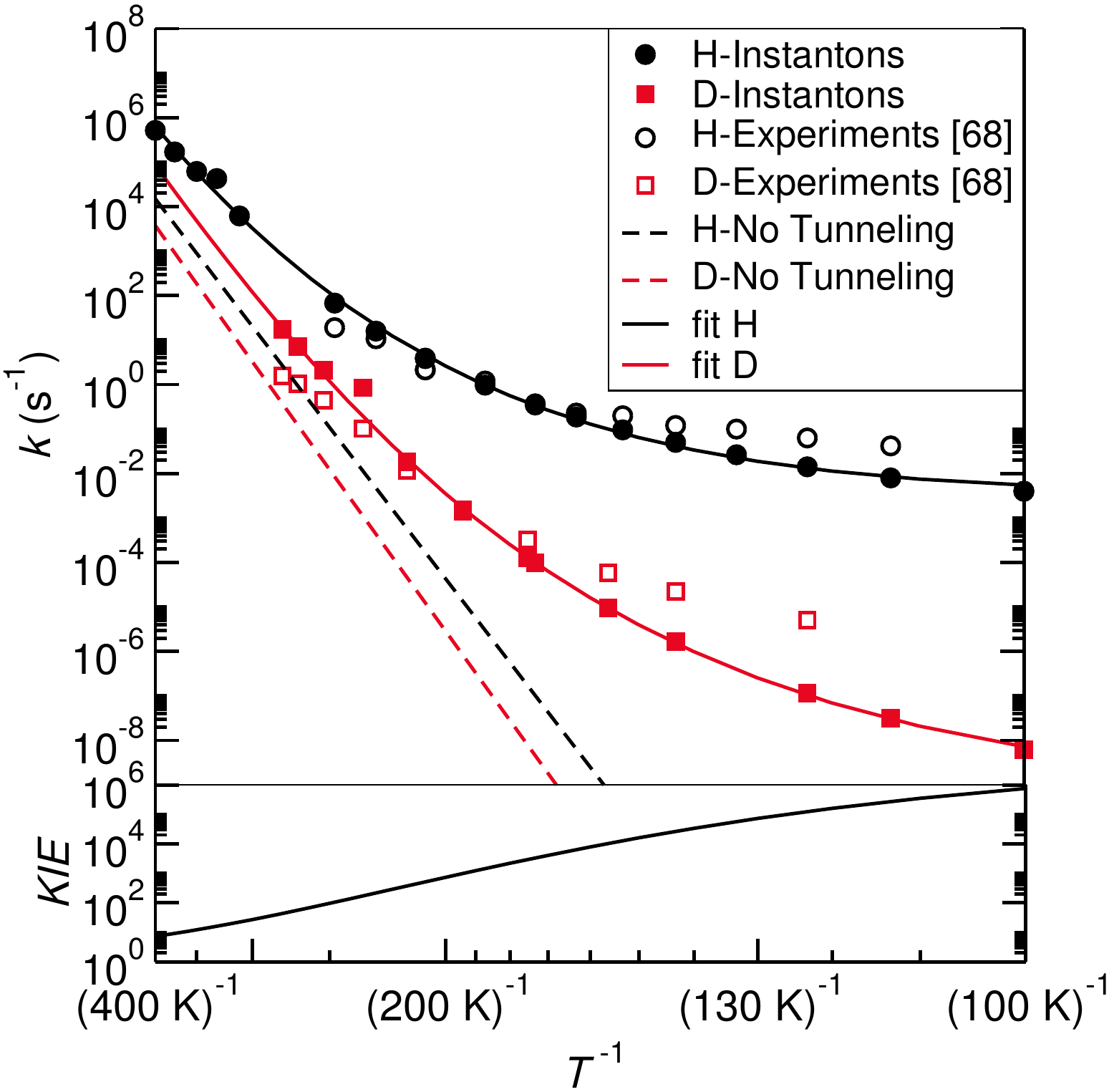}
  \caption{Arrhenius plot of rate constants obtained with the
    instanton method compared to the experimental values of reference
    \citenum{bru76} for the  H- and D-transfer reaction.
           \label{fig:instanton_rates}
          }
  \end{center}
\end{figure}

Rate constants have been calculated for temperatures down to 100~K.  Despite
the different symmetry along the IRC or instantons, the rotational symmetry
numbers $\sigma_i$ of reactant structure, transition structure and all instantons
are equal to one\cite{fer07a} and therefore, the reaction's rotational
symmetry number $\sigma=1$. Nevertheless, the rate increases by an additional
factor of $\eta =2$. For the $C_1$-symmetric instantons this is caused by
their chirality.\cite{fer07a} Even for the $C_s$-symmetric TS, $\eta =2$ must
be applied because it corresponds to the abstraction of just one specific
hydrogen atom from the methyl group which is indistinguishable from the
hydrogen atom at the other side of the mirror plane in \textbf{RS}.  At higher
temperatures, the methyl group can freely rotate and the factor might be
increased to $\eta_{\text{free}} =3$. In the Arrhenius plot in
\figref{fig:instanton_rates} we used $\eta = 2$ throughout all temperatures.

Accordingly we assume the \emph{tert-}butyl groups and the other individual
methyl groups to be hindered in their rotation, too, and treat them as
harmonic oscillators.  In general the difference of describing methyl or
\emph{tert-}butyl groups as free rotors instead of harmonic oscillators should
be minor because the effects in the reactant structure and transition structure
cancel.  Obviously, the same argumentations are valid for the perdeuterated
system.

Rate constants, including those of the perdeuterated system,
as well as rate constants without consideration of atom tunneling
(calculated by means of harmonic transition state theory)
are shown in \figref{fig:instanton_rates}.
 A full list of all values is given
in the Supplementary Information. We performed a fit of the rate constants to the equation
\cite{zhe10}
\begin{equation}
k(T) = A 
\exp \left(- \frac{E_0}{R} \frac
{T+T_0}
{T^2+T_0^2}
\right)
\end{equation}
which shows good agreement with the instanton values as well as the experimental values.
The fit of $k(T)$ allows us to calculate the KIEs, see \figref{fig:instanton_rates}.
The KIE at 100~K is almost $10^{6}$.

\begin{table}[h!]
 \caption{Fitting parameters for the hydrogen and deuterium transfer reactions.
 \label{tab:parameters}
         }
    \begin{center}
\setlength{\tabcolsep}{2mm}
\begin{tabular}{lrr}
\hline
      &       H-transfer    &    D-transfer  \\
\hline
$A$ / s$^{-1}$  & $7.45\cdot 10^{13}$   &   $5.25\cdot 10^{13}$               \\
$E_0$ / kJ mol$^{-1}$&    51.51 & 55.77\\
$T_0$ / K       &            200.2              &         154.8       \\
\hline
\end{tabular}
\end{center}
\end{table}

\subsection{Reaction of HCl + Cl}
\begin{figure}[bt]
  \includegraphics[width=8cm,clip]{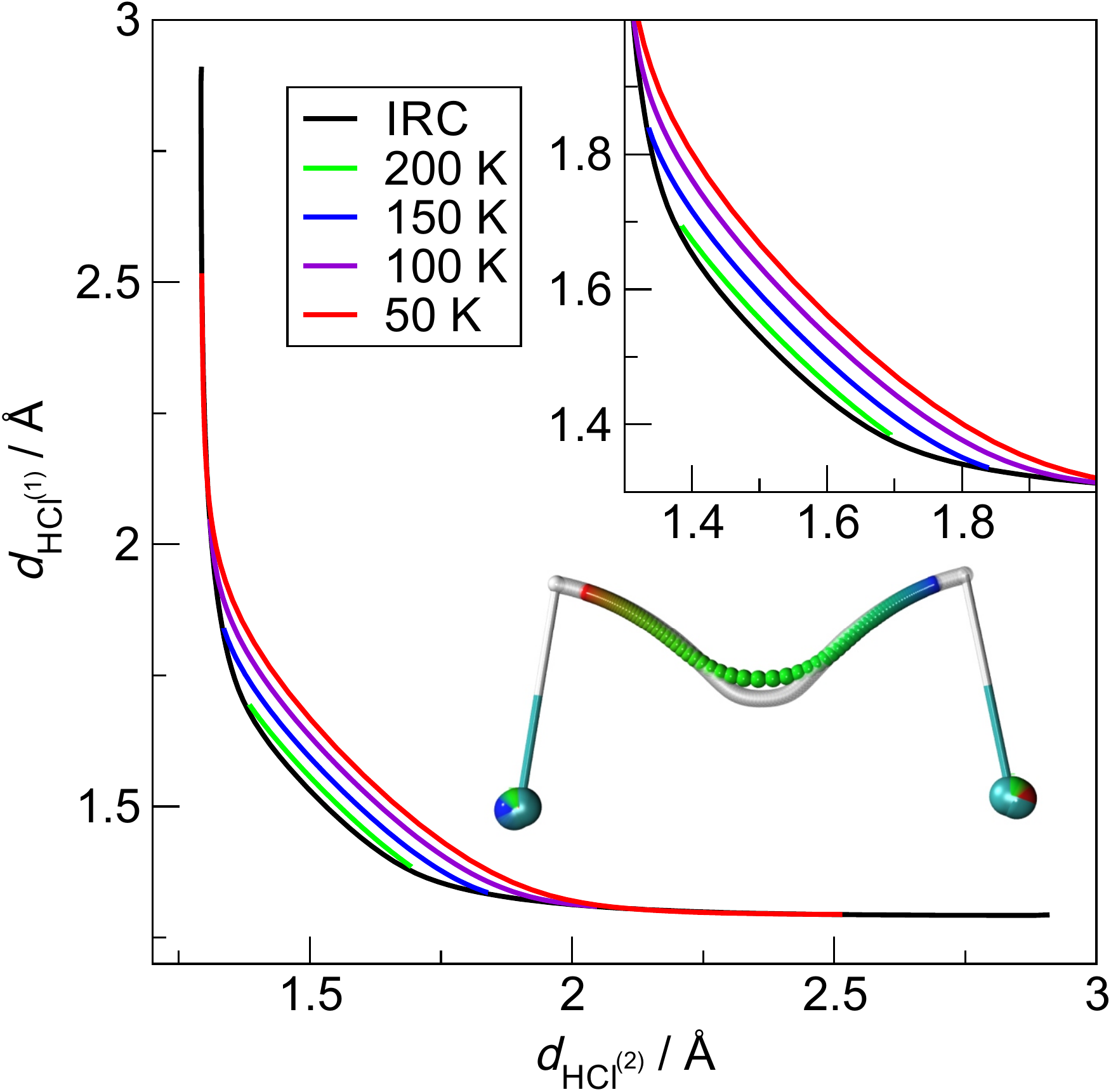}
  \caption{
\hl{
  Corner cutting in the reaction of HCl + Cl:
  The Cl--H distances are reduced during the tunneling process.
  At low temperature the corner cutting leads to a significant reduction of the total path length.
  The picture of the molecular system shows the classical IRC (white path) and the instanton at 50~K (from red to green to blue).
}
    \label{fig:ClH}
  }
\end{figure}

\begin{figure}[bt]
  \includegraphics[width=8cm,clip]{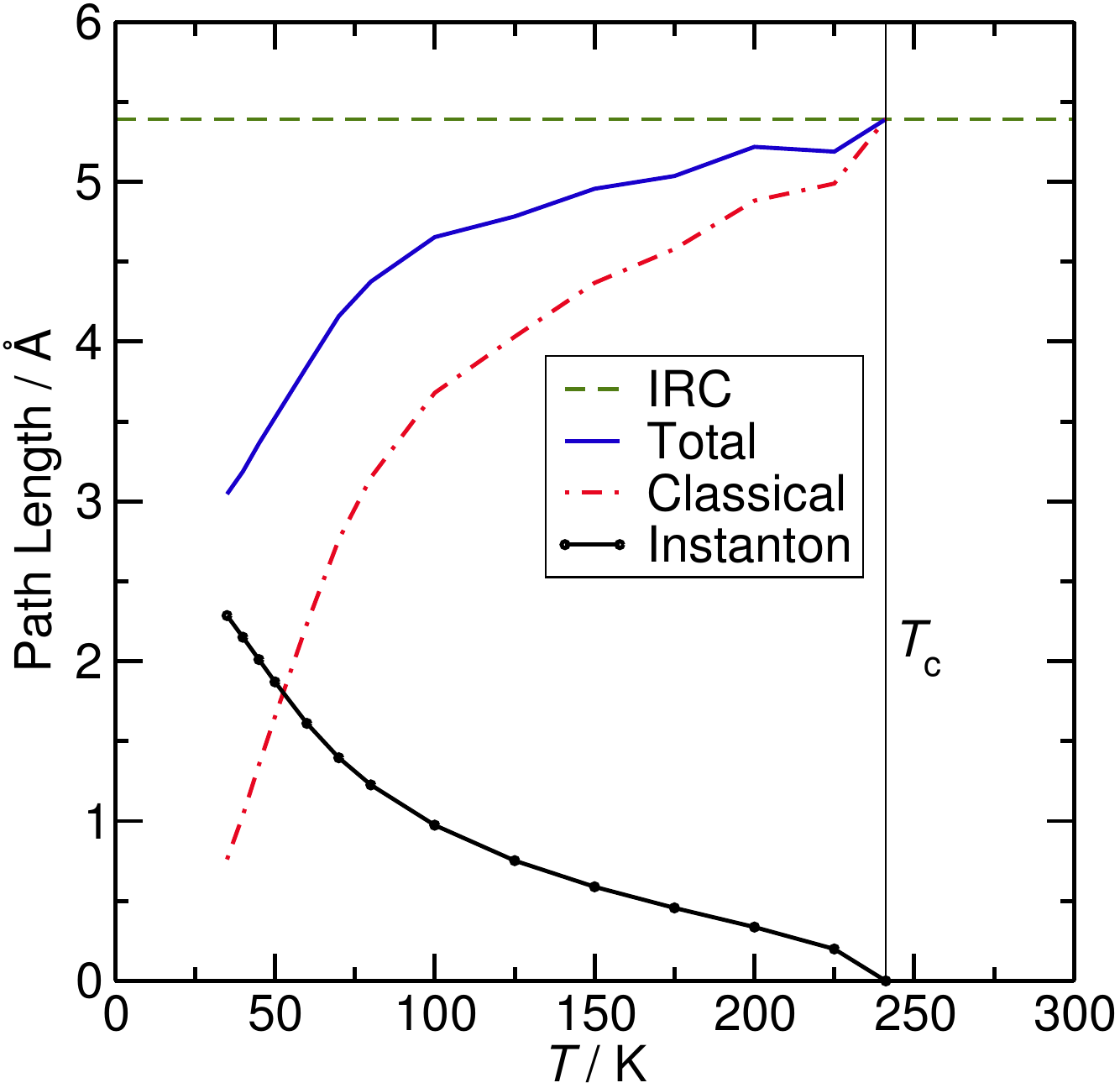}
  \caption{
\hl{
  Contributions of instantons and classical paths to the total path length at various
  temperatures. 
}
    \label{fig:ClH:length}
  }
\end{figure}

\hl{
As a complementary example, we want to present the atom transfer reaction 
}
\begin{equation}
\text{HCl} + \text{Cl}
\rightarrow  
\text{Cl} + \text{HCl}
\end{equation}
\hl{
which is an example for a symmetric 
double-well potential and a prototypic heavy--light--heavy reaction.\mbox{\cite{kneba1979,bu1996,gonzales1998,ju1990,moradi2015}}
Thermal rate constants for this reaction can not be observed experimentally
except of the reaction with isotopically labelled chlorine.
Therefore, we restrict ourselves to the presentation of the instantons and
paths.

We calculated the potential energy barrier to be 24.9~kJ~mol$^{-1}$.
The barrier is of medium width and the crossover temperature is $T_\text{c} =241.1$~K.
We calculated the potential energy along the IRC and instanton paths from 200~K to 35~K.
To elucidate the corner cutting effect, 
one Cl--H distance is plotted against the other Cl--H distance 
in \mbox{\figref{fig:ClH}}.
}
\hl{
As above, high-temperature instantons are closer to the IRC.
At lower temperatures, the instantons shorten the distance the hydrogen atom has to cover.
For this reaction, the low-temperature instantons are qualitatively of the same shape as the IRC.

The path during a reaction involving atom tunneling 
can be decomposed into the instanton and 
the classical steepest-descent paths on the potential energy hypersurface 
starting from the turning points of the instanton on both sides of the barrier.
The instanton path length, the classical path length,
and the sum of both are shown 
in \mbox{\figref{fig:ClH:length}} for different temperatures.
At temperatures next to $T_\text{c}$
the spread of the instanton is small and the classical path 
is nearly as long as the full IRC.
At lower temperature, the instanton spreads out. The classical path becomes smaller
and the sum of both contributions is significantly smaller than the IRC.
At 35~K the total tunneling path length is approximately 56.6\% of the IRC
path length
which demonstrates the pronounced corner cuttinng effect.
}

\section{Conclusions}

In this paper we compare the classical reaction path (IRC) with the tunneling
path from semiclassical instanton theory.  The instanton, which is the most
likely tunneling path at a certain temperature, can deviate qualitatively from
the classical reaction path. 
\hl{
As examples we used a reaction where a
hydrogen atom is transferred from a methyl group to a phenyl radical 
({\bf 1} $\rightarrow$ {\bf 2})
and the reaction of HCl + Cl,
which is a prototype for atom tunneling in a heavy--light--heavy arrangement.
}

\hl{For the reaction {\bf 1} $\rightarrow$ {\bf 2}
experimental} data indicated the importance of
tunneling.\cite{bru76} Instanton theory can reproduce the
experimental rate constants within a reasonable accuracy.  We performed fits
to the instanton rate constants and used these to calculate kinetic isotope
effects.

To achieve the comparison, we implemented a modified Hessian
predictor-corrector algorithm for the calculation of IRCs.  The algorithm uses
quadratic information of the potential hyper surface.  The scaling is
below $\mathcal{O}(N^3)$ because any matrix diagonalizations are avoided.
Therefore, it is also suitable for larger systems as long as a single Hessian
calculation at the TS can still be carried out.

\hl{
We have quantified the corner cutting effect by means of
a combination of instanton paths and classical paths
in the reaction of HCl + Cl.
}

\section*{Acknowledgments}
This work was financially supported by the German Research Foundation (DFG)
within the Cluster of Excellence in Simulation Technology (EXC 310/2) at the
University of Stuttgart.  The authors acknowledge support in terms of CPU time
by the state of Baden-W\"urttemberg through bwHPC and the Germany Research
Foundation (DFG) through grant no INST 40/467-1 FUGG. This work was
financially supported by the European Union's Horizon 2020 research and
innovation programme (grant agreement No. 646717, TUNNELCHEM).

\bibliography{mod} 
\bibliographystyle{rsc} 

\end{document}